\documentclass[nohyper,11pt,letterpaper]{JHEP3}
\usepackage{graphicx}
\usepackage{amsmath}
\usepackage{amssymb}
\usepackage{amsthm}
\usepackage{amsfonts}

\title{Non-Gaussianities in the Cosmological Perturbation Spectrum due to Primordial Anisotropy}

\author{Anindya Dey\\
Theory Group, Department of Physics and Texas Cosmology Center\\ The University of Texas at Austin,
TX 78712.
\\ E-mail: \email{anindya@physics.utexas.edu}}

\author{Sonia Paban\\
Theory Group, Department of Physics and Texas Cosmology Center\\ The University of Texas at Austin,
TX 78712.
\\ E-mail: \email{paban@physics.utexas.edu}}

\abstract{We investigate possible signatures of a pre-inflationary anisotropic phase in two-point and three-point correlation functions of the curvature perturbation for high-momentum modes which exit the horizon well after isotropization. In this momentum regime, the early time dynamics admits a WKB description and the late time dynamics can be described in terms of a non-Bunch Davies vacuum state which encodes the information of initial anisotropy in the background spacetime. We compute the bi-spectrum for curvature perturbation in a canonical single-field action with and without higher derivative operators. We show that the bi-spectrum at late times, in either case, is enhanced for a flattened triangle configuration as well as a squeezed triangle configuration and compute the corresponding $f_{NL}$ parameters. The angular dependence and the particular momentum dependence of the $f_{NL}$ parameter appear as distinctive features of background anisotropy at early times.}


\received{???????? ?st, 2011} \accepted{???????? ?th, 9999}
\preprint{\hepth{yymmnnn}\\UTTG-13-11\\TCC-016-11}
\newcommand{\gev}{\, \mbox{GeV}}
\newcommand{\Mpc}{\, \mbox{Mpc}}
\begin{document}

\section{\bf Introduction}

A deeper understanding of the inflationary scenario requires going beyond the power spectrum and probing higher correlation functions, collectively referred to as non-Gaussianitiy  \cite{Komatsu:2009kd}. Observable non-Gaussianity, in turn,  requires a departure from the standard single-field inflation with a canonical action \cite{Maldacena:2002vr, Creminelli:2004yq}. Substantial progress has been made in understanding the enhancement in non-Gaussianity for several variants of the standard scenario, which involve having multiple scalar fields, non-canonical action for the scalar field, introducing higher derivative terms in the action or having a non-standard vacuum state (see  \cite{Bartolo:2004if} for reviews). A common feature of all these models is that they have a homogeneous and isotropic background for the perturbations to evolve.\\
In the present work, we investigate the possible signatures of a strong anisotropy in the early space-time metric in the perturbation spectrum and its non-Gaussianity. 
The effect of primordial anisotropy in the power spectrum has been studied before by \cite{Pereira:2007yy, Gumrukcuoglu:2007bx} and most recently  by \cite{Kim:2010wra}. We redo the computation here for completeness.   A primordial anisotropy in the metric will be washed away by a period of inflation, hence it seems interesting to try to find signatures that do not get washed away. 

We focus on a family of axially symmetric Bianchi I background geometries which admit a WKB solution for the perturbations at early times for modes in the high-momentum regime. On matching the WKB solution with the solution at late times, we can describe the late time dynamics of the curvature perturbation in terms of a non-standard ground state (essentially an excited state on the BD vacuum). The possible enhancement of the non-Gaussianity for non-standard (non Bunch-Davis) vacuum was pointed out years ago by Holman and Tolley  \cite{Holman:2007na}.  By studying the three-point correlation function in the present scenario and deriving the relevant contribution of background anisotropy to the $f_{NL}$ parameter, we infer that a possible enhancement in the bi-spectrum may also occur in the squeezed triangle limit, in addition to the
flattened triangle limit discussed in \cite{Holman:2007na}.\\
Although we are primarily interested in non-canonical vacuum states which arise as a result of spatial anisotropy at early times , our computation easily generalizes to any excited state obtained by a Bogoliubov 
transformation on the Bunch-Davis vacuum.\\
The paper is organized as follows. In the next section, we discuss the anisotropic background geometry in which the inflaton evolves and study the classical equations of motion. Section 3 deals with the study of cosmological perturbations in the high-momentum regime and the WKB solution. In section 4, we compute the bi-spectrum first for a canonical action and then in presence of higher derivative terms.\\
While we were preparing this manuscript, we came across the work \cite{Kim:2011pt}, which has some overlap with our work.\\

\section{\bf Background Equations of Motion}
We consider a theory of Einstein gravity with a minimally coupled single scalar field given by the following action,
\begin{equation}
S=\frac{1}{2}\int d^4x \sqrt{-g} R + \int d^4x \sqrt{-g}\left(-\frac{1}{2}g^{\mu\nu}\partial_{\mu}\phi\partial_{\nu}\phi -V(\phi)\right), \hspace{3ex}(M_p^2\equiv 1)
\end{equation}
where the background metric is chosen to be an axially symmetric version of the
Bianchi I metric:
\begin{equation}
ds^2=-dt^2 + e^{2\rho}(dx^1)^2 +e^{2\beta}(dx^{\alpha})^2 \label{metric}
\end{equation}
with $\alpha=2,3$. \\
In contrast to the FRW case where one has a single Hubble constant, we have two Hubble constants, which we choose to define as follows:
\begin{equation}
H=\frac{\dot{\rho}+2\dot{\beta}}{3}, \hspace{2ex} h=\frac{\dot{\rho}-\dot{\beta}}{\sqrt{3}}
\end{equation}
The classical dynamics of the system specified by the action (2.1) constitutes a strongly anisotropic expansion at early times (parametrized by $h$) followed by eventual isotropization at a time-scale $t \approx t_{iso}=\frac{1}{\sqrt{V}}$. For $t \gg t_{iso}$, the universe enters a phase of de Sitter expansion.\\

Note that $h$, which, roughly speaking, is a measure of the rate of anisotropic expansion vanishes in the isotropic limit ($\dot{\rho}=\dot{\beta}$) so that we are left with a single Hubble constant.\\

In terms of $H$ and $h$, the independent Einstein's equation and the equation of motion for the scalar field reduce to the following set of equations:
\begin{eqnarray}
\dot{H}+3H^2 &=&V(\phi)\\
3H^2-h^2 &=& \frac{1}{2}\dot{\phi}^2 + V(\phi)\\
\ddot{\phi}+3H\dot{\phi}+V'(\phi)& = & 0
\end{eqnarray}

The time-evolution of $h$ can easily be derived from the above equations,
\begin{equation}
h(\dot{h}+3Hh)=0
\end{equation}
In the anisotropic phase, $h \neq 0$, which leads to the equation of motion,
\begin{equation}
\dot{h}+3Hh=0
\end{equation}

For a general $V(\phi)$, one can only obtain approximate solutions to the above system of equations. However, for a constant $V$, one can exactly solve the coupled differential equations for $H,h$ and $\dot{\phi}$ as follows,
\begin{eqnarray}
H& =& \sqrt{\frac{V}{3}} \, \coth({\sqrt{3V}t}) = H_I \,  \coth({\sqrt{3V}t}) \nonumber\\
h & = & \pm \sqrt{V} \, \frac{1}{\sinh({\sqrt{3V}t})}  \label{sip}\\
\dot{\phi} &= & 0  \nonumber
\end{eqnarray}

In the above solution, the constants have been chosen such that the metric approaches a Kasner solution in the limit $t \rightarrow 0^+$. The $\pm$ sign in the solution of $h$ indicates two different branches in the solution space (distinguished, among other things, by their behavior in the Kasner limit). It turns out that only for the positive branch, one can impose initial conditions on the cosmological perturbations at early times via the usual WKB approximation \cite{Gumrukcuoglu:2008gi}. Hence, in this note, we will focus exclusively on this class of backgrounds.\\

Now, for a given non-trivial $V(\phi)$, the slow-roll condition ($\ddot{\phi}
\approx 0$) will imply that at early times $V(\phi)$ is nearly constant with time, provided $H\dot{\phi}^2 \rightarrow 0$ at early times. This condition is obeyed by all common inflaton potentials and hence the above solution (\ref{sip}) can be trusted for a non-constant potential in the $t \rightarrow 0^+$ limit. As an example, consider $V(\phi)=\frac{m^2\phi^2}{2}$ for which $H$ and $\phi$ have the following asymptotic forms at early times,
\begin{eqnarray}
H & = & \frac{1}{3t}\left[1+\frac{m^2\phi_0^2t^2}{2}+O(m^4t^4)\right]\\
\phi & = & \phi_0 \left[1-\frac{m^2t^2}{4}+O(m^4t^4) \right] 
\end{eqnarray}
In this case, $H\dot{\phi}^2 \approx t \rightarrow 0$, so that $V$ is essentially constant at early times.\\

In this Kasner limit, the metric reduces to the following form,
\begin{equation}
ds^2_{Kasner}=-dt^2 + (\sqrt{V}t)^2 (dx^1)^2 + (dx^{\alpha})^2
\end{equation}
with $\dot{\rho}=\frac{1}{t}$, $\dot{\beta}=0$.\\

This is the gravitational background in which the cosmological perturbations
evolve at early times. The solutions for the background equations of motion suggest that the universe starts life with a very strong anisotropy ($h\rightarrow \frac{1}{t}$ at early times) which is smoothed out very fast by the inflaton potential. The universe then enters a phase of usual isotropic inflation.\\

\section{Cosmological Perturbations: Small Wavelength Limit}
The computation of the spectrum for cosmological perturbations for a generic anisotropic background has two significant differences with the corresponding computation in the isotropic case:\\
(1) The existence of a WKB solution for the modes of a given wavelength
at asymptotically early times ($t\rightarrow 0^+$) is not guaranteed , since 
for certain backgrounds any mode may become super-Hubble \cite{Pereira:2007yy} in this limit.\\
(2) The $SO(3)$ scalar and tensor perturbations are coupled for a generic wavelength at times $t \leq t_{iso}$ \cite{Gumrukcuoglu:2007bx}.\\

As commented in the previous subsection, the particular choice of the ``positive branch'' background solves (1). The positive branch metric in the Kasner limit is, in fact, a patch of Minkowski space-time \cite{Kim:2010wra} and this can be seen as follows:
Let $u=t \, \sinh({Vx^1})$ and $v=t \, \cosh({Vx^1})$, so that,
\begin{equation}
ds^2=-dt^2 + (\sqrt{V}t)^2 (dx^1)^2 + (dx^{\alpha})^2=-dv^2 + du^2 + (dx^{\alpha})^2
\label{metricMin}
\end{equation}
Therefore, this particular background admits a WKB solution for perturbations at early times. Note that  coordinate invariants such as the curvature and the Weyl tensor are time independent and non vanishing and hence the space is never Minkowski.  Thus, when computing quantities that depend on derivatives of the metric one should be careful if using (\ref{metricMin}).

In \cite{Gumrukcuoglu:2007bx}, the perturbations in the anisotropic phase were parametrized in terms of the variables $v,H_{\times},H_{+}$ (which reduce to the usual gauge-invariant variables in the isotropic limit, $v$ becomes the Mukhanov variable and $H_{\times},H_{+}$ become the two polarizations of the tensor modes) with the following equations of motion:
\begin{eqnarray}
H_{\times}'' +\omega^2_{\times}H_{\times} & =& 0\\
\begin{pmatrix} v\\H_+ \end{pmatrix} ^{''} & =& \begin{pmatrix} \omega^2_{11}& \omega^2_{12}\\\omega^2_{21}&\omega^2_{22} \end{pmatrix} \begin{pmatrix} v\\H_{+}  \end{pmatrix}
\end{eqnarray} 
where the derivative is with respect to the conformal time $\eta$ and the 
frequencies of the coupled system are given as,
\begin{eqnarray}
\omega_{11}^2 & = & e^{2\rho} \, (p_1^2+p_2^2- 2\dot{\rho}\dot{\beta}+\cdots)\\
\omega_{22}^2 & = & e^{2\rho}\, (p_1^2+p_2^2- 2\dot{\rho}\dot{\beta}+\cdots)\\
\omega_{12}^2 & = & e^{2\rho}\,  \left(\frac{\sqrt{2}p_2^2(\dot{\rho}-\dot{\beta})}{2\dot{\beta}p_1^2+(\dot{\rho}+\dot{\beta})p_2^2} \right) \, \left(-\frac{-3\dot{\beta}\dot{\phi}}{M_p}+\cdots \right)
\end{eqnarray}
The $p_i$ are the physical momenta, $p_1= k_1 e^{-2 \rho}, p_{\alpha}= k_{\alpha} e^{-2 \beta}$. The ellipsis in the above equations indicates terms subleading in the limit $t\rightarrow 0^+$. To the leading order at asymptotically early times, $\omega_{11}^2 \, e^{-2\rho},\omega_{22}^2\, e^{-2\rho} \approx 1/t^2$ while $\omega_{12}^2\, e^{-2\rho} \approx t^2$  and as a result, the mixing terms can be neglected. Therefore, in this limit, the scalar and tensor perturbations decouple as in the isotropic case. This feature is not surprising since, as seen earlier, the metric is that of a flat space-time.\\

In this work, we focus on the fluctuations  of the scalar mode (curvature perturbation)- the computation for tensor perturbation can be done similarly. In the isotropic case, the equation of motion of the Mukhanov variable $v$ is identical to that of  a scalar field evolving in the same background, as long as the slow-roll conditions are obeyed (as a result of which $\frac{z''}{z} \approx \frac{a''}{a}$). Therefore, in all quantities of interest, one can substitute curvature perturbation by a solution of the  scalar field equation of motion, up to a well-defined normalization. In the anisotropic case, the scalar mode is given as 
$v=\exp{2\beta}[\delta\phi + \frac{p_2^2\dot{\phi}}{\dot{\rho}p_2^2+\dot{\beta}(2p_1^2+p_2^2)}\psi]$ in terms of the scalar field and metric fluctuation \cite{Gumrukcuoglu:2007bx}. Since $v$ approaches the usual Mukhanov variable in the isotropic limit, the above argument is true for the scalar mode evolving in the anisotropic background in the limit $t \gg t_{iso}$ when the universe enters a late-time de Sitter phase.\\
At asymptotically early times, $v$  reduces to purely a fluctuation in the inflaton field in the Kasner background. In the limit $t \rightarrow 0^+ $,  $\frac{p_2^2\dot{\phi}}{\dot{\rho}p_2^2+\dot{\beta}(2p_1^2+p_2^2)} \approx t^2$, so that $v \approx \delta\phi$.\\
Thus, both at early and late times, the curvature perturbation can be understood as a scalar field evolving in the background given by the metric (2.2). Therefore, if one can find a WKB solution at early times, one can obtain an approximate classical solution  at late times by a standard matching procedure at some intermediate time. This approximate classical solution will specify the particular vacuum state of the curvature perturbation field and can then be used to compute the late-time correlation functions. It is important to note that, in this scheme, the entire information of anisotropy is encoded in the vacuum state of the theory.\\

Therefore, we consider a scalar field propagating in the background metric (\ref{metric}). It turns out that (as discussed in Appendix A) there exists a WKB solution for the scalar field in the Kasner regime for $k \gg H_I$, where $H_I=\sqrt{\frac{V}{3}}$ and $V=V(\phi)]_{t\rightarrow 0}$. As explained in the appendix, the WKB solution amounts to imposing a particular initial condition on the scalar modes at early times. In this work, we will only be concerned with the non-planar regime, viz. $k_1 \approx k_2 \sim k_3$, where the WKB condition is always satisfied at early times. Note that the condition $k \gg H_I$ is equivalent to the condition of the modes being deep inside the horizon at early times, $\frac{k}{e^{\rho}\dot{\rho}} \gg 1$ with $e^{\rho}\approx \sqrt{V}t$ and $\dot{\rho}=\frac{1}{t}$.
For the observed wavelength scales, between $(1-10^4) \Mpc$, it can be shown that \cite{Kim:2010wra} 

\begin{equation} e^{N-64} \left( \frac{T_R}{10^{14} \gev }\right) \left( \frac{10^{16} \gev}{V^{1/4}}\right)^2  \,\, < \frac{k_{obs}}{H_I} \,\, <  e^{N-55} \left( \frac{T_R}{10^{14} \gev }\right) \left( \frac{10^{16} \gev}{V^{1/4}}\right)^2 \label{bounds} \end{equation} where $N$ is the number of e-foldings and $T_R$ is the reheating temperature. For the range of observable scales satisfying the WKB condition is not hard. The visibility of the anisotropy at these scales, however, is not guaranteed. 

\begin{table}[htdp]
\caption{Relations between different time scales}
\begin{center} 
\begin{tabular}{|c|c|c|c|} \hline
Time & Definition & Relations & $e^{\alpha(t)}$ \\   \hline 
$t_{iso}$ &  $\rho(t_{iso}) \sim \beta(t_{iso}) \sim \alpha(t_{iso})$ &&$ O(10^0)$ \\ \hline
$t_{*}$ & $e^{\alpha(t_*)} \equiv \sqrt{\frac{k}{H_I}} $ & $t_* > 5 t_{iso}$ &   $  >  O(10^2)$ \\ \hline
$t_e$ & $e^{\alpha(t_e)} \equiv \frac{k}{H_I}$ & $t_e  \sim 2 t_{*} $&  $  >  O(10^4)$ \\ \hline
\end{tabular}
\end{center}
\label{default}
\end{table}

At asymptotically late times \cite{Kim:2010wra}, there exists a general solution whose precise form can be determined by matching the early time WKB solution at some intermediate time $t_{*}$.\\
The solution for the scalar field at late times can then be written as,
\begin{equation}
\phi=A_{+}({\bf k})\phi_{+}(\eta)+A_{-}({\bf k})\phi_{-}(\eta) 
\end{equation}
where $\phi_{\pm}(\eta)=(1\mp ik\eta)\exp{(\pm ik\eta)}$, with $\eta$ being the usual conformal time as defined in a de Sitter universe.\\
The coefficients $A_{+}$ and $ A_{-}$  are given as,
\begin{eqnarray}
A_{+} & =& \frac{i\varepsilon^3}{2\sqrt{2H_I}}\left[(2-\varepsilon^2)+2i\varepsilon\left(\frac{\varepsilon^2}{2}-1\right) +O(\varepsilon^4)\right]\exp{\left(\frac{-i}{\varepsilon}\right)} \nonumber\\ \nonumber\\
A_{-} & = & \frac{i\varepsilon^3}{2\sqrt{2H_I}}\left[\left(\frac{2}{3}-r^2 \right)\varepsilon^3 +O(\varepsilon^4) \right]\exp{\left(\frac{i}{\varepsilon}\right)} \label{apam}
\end{eqnarray}
where $\varepsilon=\sqrt{\frac{H_I}{k}}$ and $r=\frac{\sqrt{|k_y^2+k_z^2|}}{k}$.\\
The WKB approximation is valid when $\varepsilon \ll 1$  and we have retained terms up to order $\varepsilon^3$, which is the minimal order at which any signature of anisotropy appears.

From the above solution, it follows that,
\begin{equation}
|\phi|^2_{\eta \rightarrow 0} \longrightarrow |A_{+}+A_{-}|^2=\frac{H_I^2}{2k^3}\left[1+
\left(\frac{2}{3}-r^2 \right) \left(\frac{H_I}{k}\right)^{3/2} \cos\left(2\sqrt{\frac{k}{H_I}}\right)\right]
\end{equation}
Background anisotropy, therefore, implies that the ground state for the curvature perturbation field at late times is not given by the usual Bunch-Davies vacuum but an excited state built on the Bunch-Davies vacuum.

From this, the late time correlation and power spectrum for the curvature perturbation can be easily derived. We need to replace the parameter $H_I$ by the Hubble parameter of the universe at horizon exit $\dot{\rho}_e$ and introduce the overall normalization factor of $\frac{\dot{\rho}}{\dot{\phi}}$ also evaluated at horizon exit.\\
Therefore, the two-point correlation function for the curvature perturbation is given as,\\
\begin{eqnarray*}
\left\langle \zeta_{\bf{k}}(t)\zeta_{\bf{k'}}(t)\right\rangle & \approx & (2\pi)^3\delta^3({\bf k}+{\bf k'}) \, \frac{\dot{\rho_e}^2}{2k^3} \, \frac{\dot{\rho}_e^2} {\dot{\phi}_e^2}\left[1+ \left(\frac{2}{3}-r^2\right)\left(\frac{\dot{\rho_e}}{k}\right)^{3/2} \cos\left(2\sqrt{\frac{k}{\dot{\rho_e}}}\right) \right] \\ \\
& \equiv & (2\pi)^3\delta^3({\bf k}+{\bf k'}) \, \frac{F(k,\cos{\theta})}{2k^3}
\end{eqnarray*}
where 
\begin{eqnarray}
\dot{\rho}(t_e) \, e^{\rho(t_e)} &\approx & k\\
\cos{\theta} &= & \frac{k_x}{k}= \sqrt{1-r^2} 
\end{eqnarray}

The spectral index for the curvature perturbation is then given as,
\begin{eqnarray}
n_s-1=k\frac{d}{dk}\log[F(k,\cos{\theta})] & \approx &   \frac{1}{\dot{\rho}_e} \frac{d}{d t_e}\log[F(k,\cos{\theta})] \nonumber \\ & \approx &  2(\eta -3\epsilon) 
 + \left(\frac{1}{3}-\cos^2{\theta}\right)e^{-\rho_e}\sin{(2 e^{\rho_e/2})} + O(e^{-3 \rho_e/2}) \nonumber \\
\end{eqnarray}
Note that, $e^{-\rho_e} \sim e^{-t_e/t_{iso}}$. One can easily estimate the magnitude of the correction term arising purely due to early-time anisotropy.
In the appendix, we show that the time $t_*$ (time at which we match the WKB solution with the de Sitter solution) obeys $t_* > t_{iso}$, such that $e^{\rho(t_*)} \gg 1$. In addition, we show that $t_e \approx 2 t_*$. Therefore, for $t_* = 5 t_{iso}$, for example, we have $t_e = 10 t_{iso}$, which implies that the correction term is of the order $e^{-\rho_e} \sim e^{-t_e/t_{iso}} = e^{-10} \sim 10^{-5}$, while the slow-roll terms are of the order $10^{-2}$. In this case, we are looking at a regime of momenta where $\frac{k}{H_e} \approx 10^5$ or, $k \approx 10^{-1} M_p$ ($H_e \sim 10^{-6} M_p$).\\  This computation leads us to conclude that because our analytical results are valid for modes that exit the horizon well after the universe has isotropized, the effect of the anisotropy is severely suppressed in the two point function. It is important to note,  however, that there is a corner of parameter space (\ref{bounds})
(for which this analytical calculation is valid) such that the contribution of anisotropy to $n_s$ is large enough. For $t_* = 2 t_{iso}$, we have $t_e = 4 t_{iso}$
which gives a correction term of the order of $e^{-t_e/t_{iso}} = e^{-4} \sim 1/50$, comparable to the slow-roll terms. In this case,  $\frac{k}{H_e} \approx 50$, such that the WKB condition is still obeyed. We will, however, be interested in the momentum regime where the contribution of anisotropy to the two-point function is negligible and investigate its possible observable signature in the three-point function.\\

%For $t_e/t_{iso} \sim O(10)$, the angular dependence of the spectral index, arising from the background anisotropy is exponentially suppressed by the scale factor at the horizon exit of the mode.\\ 
It is interesting to compare the two-point function obtained above with the ACW parametrization \cite{Ackerman:2007nb} of the power-spectrum in a generic model of inflation with broken rotational invariance.
In \cite{Ackerman:2007nb}, the power spectrum was parametrized as 
\begin{equation}
P(k)=P(k)_0 (1+ g(k) (\hat{k}.\vec{n})^2) \label{acw}
\end{equation}
where $P(k)_0$ is the usual nearly scale-invariant contribution while $\vec{n}$ is an unit vector in a direction which breaks the rotational invariance. The power spectrum we have obtained is a slightly general form
of  $(\ref{acw})$ and admits the parametrization, 
\begin{equation}
P(k)=P(k)_0 (1+f(k)+g(k)(\hat{k}.\vec{n})^2)
\end{equation} 
where $f(k)=-\frac{1}{3}\left(\frac{\dot{\rho_e}}{k}\right)^{3/2} \cos\left(2\sqrt{\frac{k}{\dot{\rho_e}}}\right) $ and $g(k)=\left(\frac{\dot{\rho_e}}{k}\right)^{3/2} \cos\left(2\sqrt{\frac{k}{\dot{\rho_e}}}\right)$. The 
direction $\vec{n}$, in this case, can be identified with the x-axis (the scale factor along which differs from that in the axially symmetric orthogonal space), so that $\cos{\theta} \equiv \hat{k}.\vec{n}$.\\

In the next section, we compute the three point correlation function for curvature perturbations in the ground state described above and analyze the effect of background anisotropy on non-Gaussianity of the spectrum at this level.\\

\section{Computation of the 3-point function}
For computing the 3-point function for the curvature perturbation, following  \cite{Maldacena:2002vr}, we consider a local (in time) non-linear field redefinition of $\zeta$:
\begin{equation}
\zeta=\zeta_c +\frac{\ddot{\phi}}{2\dot{\phi}\dot{\rho}}\zeta_c^2 + \frac{\dot{\phi}^2}{8\dot{\rho}^2}\zeta_c^2 + \frac{\dot{\phi}^2}{4\dot{\rho}^2}\partial^{-2}(\zeta_c\partial^2\zeta_c)
\end{equation}
Evidently, this redefinition does not change the quadratic action which implies that $\zeta_c$ and $\zeta$ have the same equation of motion and hence the same classical solution given by equation (3.1). Also, since local redefinitions do not yield any enhancement of the 3-point function, it is sufficient to compute the correlation function in terms of the redefined field $\zeta_c$. In terms of $\zeta_c$, the leading term (in slow-roll parameter) in the interaction Hamiltonian will be given as,
\begin{equation}
{\cal H}_I=-\int d^3x \, d\eta \, e^{3\rho} \left(\frac{\dot{\phi}}{\dot{\rho}}\right)^4  \dot{\rho} \,
\zeta_c^{'2}\partial^{-2}\zeta'_c
\end{equation}
where the prime denotes derivative w.r.t. the conformal time $\eta$ (defined in the de Sitter phase), and the partial indicated space derivatives. \\
We can now use the ``in-in'' formalism to compute the tree-level contributions to the 3-point function. Since there is only one kind of interaction vertex, there are only two distinct Feynman diagrams at the tree-level, viz. one with a ``right'' vertex and the other one with a ``left''( recall $\left\langle Q(t) \right\rangle_{in-in} =\left\langle [\bar{T}\exp(i\int^t_{t_0}H_I(t)dt)] Q^I(t)[T\exp(-i\int^t_{t_0}H_I(t)dt)] \right\rangle$, where $T$ and $\bar{T}$ denotes the time-ordered and the anti-time-ordered product of operators. One needs to distinguish between vertices arising out of the time-ordered product from those coming from the anti-time-ordered product and we refer to them as ``right'' and ``left'' vertices respectively).\\

Therefore, using the usual Feynman rules in the momentum space, the three-point correlation function at a conformal time $\eta$ is given as,
\begin{equation}
\left\langle \zeta_c({\bf k_1},\eta)\zeta_c({\bf k_2},\eta)\zeta_c({\bf k_3},\eta)\right\rangle_{R/L}\approx \delta^{(3)}({\bf k_1}+ {\bf k_2}+{\bf k_3})A^{R/L}({\bf k_1},{\bf k_2}, {\bf k_3},\eta)
\end{equation}
where $A^R$ and $A^L$ are given as,
\begin{eqnarray}
A^R({\bf k_1},{\bf k_2},{\bf k_3},\eta)=i\int ^{\eta}_{\eta_0}d\eta'e^{3\rho(\eta')}\left(\frac{\dot{\phi}}{\dot{\rho}}\right)^4\dot{\rho}\, \left(\sum^3_{i=1}\frac{1}{k_i^2}\right) \, \prod^{3}_{i=1}\partial_{\eta'}G_{{\bf k_i}}(\eta,\eta')\\
A^L({\bf k_1},{\bf k_2},{\bf k_3},\eta)=(A^R({\bf k_1},{\bf k_2},{\bf k_3},\eta))^{*}
\end{eqnarray}
where $G_{{\bf k_i}}(\eta,\eta')=\zeta_{cl}({\bf k_i},\eta)\zeta^{*}_{cl}({\bf k_i},\eta')$, with $\zeta_{cl}({\bf k_i},\eta)$ being the classical solution for curvature perturbation. $A^R({\bf k_1},{\bf k_2},{\bf k_3},\eta)$ denotes the contribution to the 3-point function corresponding to the tree-level Feynman diagram with a ``right'' vertex while $A^R({\bf k_1},{\bf k_2},{\bf k_3},\eta)$ denotes the contribution corresponding to the tree-level Feynman diagram with a ``left'' vertex. The final result for the 3-point correlation function of the curvature perturbation is given as,
\begin{equation}
\left\langle \zeta_c({\bf k_1},\eta)\zeta_c({\bf k_2},\eta)\zeta_c({\bf k_3},\eta)\right\rangle \approx \delta^{(3)}({\bf k_1}+ {\bf k_2}+{\bf k_3})[A^{R}({\bf k_1},{\bf k_2}, {\bf k_3},\eta) + A^{L}({\bf k_1},{\bf k_2}, {\bf k_3},\eta)]
\end{equation}
In the definition of $A^R({\bf k_1},{\bf k_2},{\bf k_3},\eta)$, one needs to make a choice of $\eta_0$ - which in standard inflationary scenario is taken to be $-\infty$. However, we will take $\eta_0$ to be of the order of the isotropization time-scale  when the universe has entered an essentially de Sitter phase, thus $\eta_0$ is near the onset of inflation and parallels the choice made in \cite{Holman:2007na} . In the computation of late-time correlation functions, the parameters $\dot{\rho}$ and $\dot{\phi}$  can therefore be assigned their respective de-Sitter values, which remain nearly constant during inflation.\\
Since we are interested in the late-time correlation functions, we can set $\eta=0$, so that,
\begin{eqnarray}
G_{{\bf k_i}}(\eta=0,\eta')& = & \frac{\dot{\rho}^2}{\dot{\phi}^2} \left( |A_{+}|^2(1+ik_i\eta')e^{-ik_i\eta'}+|A_{-}|^2(1-ik_i\eta')e^{ik_i\eta'} + \right. \nonumber\\
& &  \left.  A_{+}A_{-}^{\ast}(1-ik_i\eta')e^{ik_i\eta'}+A_{-}A_{+}^{\ast}(1+ik_i\eta')e^{-ik_i\eta'}\right)
\end{eqnarray}
\begin{equation}
\partial_{\eta'}G_{\vec{k_i}}(\eta=0,\eta')=-\frac{\dot{\rho}^2}{\dot{\phi}^2}\frac{k_i^2}{\dot{\rho}e^{\rho(\eta')}}\left[(|A_{+}|^2+A_{-}A_{+}^{\ast})e^{-ik_i\eta'}+(|A_{-}|^2+A_{+}A_{-}^{\ast})e^{ik_i\eta'}\right]
\end{equation}
In the last equation we have used $\eta' \approx -\frac{1}{\dot{\rho}\exp{\rho(\eta')}}$, which is valid in the de Sitter phase of expansion. Since we have chosen $\eta_0 \approx \eta_{iso}$, this is a good approximation for $\eta_0 < \eta^{'} < 0$.\\

Now, plugging the above expression for $\partial_{\eta'}G_{\vec{k_i}}(\eta=0,\eta')$ in equation (4.4), we have,
\begin{equation}
A^R({\bf k_1},{\bf k_2},{\bf k_3})=-i\frac{\sum_{i<j}k^2_ik^2_j}{\dot{\phi}^2}
\int^0_{\eta_0}d\eta'\sum_{{\xi_i}=\pm1}\prod^3_{i=1}e^{i(\xi_ik_i)\eta'}  F_{\xi_i}(k_i)
\end{equation}
where the sum extends over all 8 possible linear combinations $\xi_ik_i$ and
$F_{\xi_i=-1}(k_i)=|A^{i}_{+}|^2+A^{i}_{-}{A^{i}_{+}}^{\ast}$ and $F_{\xi_i=1}(k_i)=|A^{i}_{-}|^{2}+A^{i}_{+}{A^{i}_{-}}^{\ast}$.\\
Therefore, on completing the $\eta'$ integration, we have,
\begin{equation}
A^R({\bf k_1},{\bf k_2},{\bf k_3})=-\frac{\sum_{i<j}k^2_ik^2_j}{\dot{\phi}^2}
\sum_{{\xi_i}=\pm1}(\prod^3_{i=1} F_{\xi_i}(k_i))\frac{1}{\sum_i\xi_ik_i}\left(1-e^{i \eta_0\sum_i\xi_ik_i}\right)  \label{aRf}
\end{equation}

Now from the expression for $A_{\pm}$, one finds,
\begin{eqnarray}
F_{\xi_i=1}(k_i)=\frac{\varepsilon_i^6}{4\dot{\rho}}\left(-\frac{1}{3}+\cos^2{\theta_i}\right)\varepsilon_i^3 \exp\left({\frac{-2i}{\varepsilon_i}}\right)\\
F_{\xi_i=-1}(k_i)=\frac{\varepsilon_i^6}{2\dot{\rho}}\left[1+\frac{1}{2}\left(-\frac{1}{3}+\cos^2{\theta_i}\right)\varepsilon_i^3 \exp\left({\frac{2i}{\varepsilon_i}}\right)\right]
\end{eqnarray}
The leading order term in $\varepsilon_i$ is identical to what one gets in a standard computation of the bi-spectrum using the BD vacuum. The subleading term ($\approx \varepsilon^9_i$) carries the signature of background anisotropy. The leading term in $A^R$ arising from the primordial anisotropy (of the order $\varepsilon^9$) is given by the configuration ${\xi_1=1,\xi_2=-1,\xi_3=-1}$ and its permutations.\\ 
As is evident from equation (\ref{aRf}), the bispectrum can be  enhanced (in contrast to the standard case \cite{Maldacena:2002vr}) if the denominator $\sum_i\xi_ik_i$ vanishes. However, the expression does not blow up since the exponential factor in equation (\ref{aRf}) also goes to zero ( $\frac{1}{\sum_i\xi_ik_i}(1-\exp{i(\sum_i\xi_ik_i)\eta_0})\approx -i\eta_0$, in the limit $\sum_i\xi_ik_i \rightarrow 0$).\\
For the aforementioned choice of the ${\xi_i}$,i.e. $k_1=k_2 + k_3$
we have the following contribution to the bi-spectrum,
\begin{eqnarray}
\Delta\left\langle \zeta_c({\bf k}_1,0)\zeta_c({\bf k}_2,0)\zeta_c({\bf k}_3,0)\right\rangle & \equiv & \left\langle \zeta_c({\bf k}_1,0)\zeta_c({\bf k}_2,0)\zeta_c({\bf k}_3,0)\right\rangle-\left\langle \zeta_c({\bf k}_1,0)\zeta_c({\bf k}_2,0)\zeta_c({\bf k}_3,0)\right\rangle_{Isotropic}  \nonumber \\ & = & \delta^{(3)}(\sum_i {\bf k}_i) \frac{\sum_{i<j}k^2_ik^2_j}{\dot{\phi}^2}\frac{1}{16\dot{\rho}^3}\prod^3_{i=1}\varepsilon_i^6 \left[ \left(-\frac{1}{3}+\cos^2{\theta_1}\right)\varepsilon_1^3 \sin{\frac{2}{\varepsilon_1}}\right] \, \eta_0 \nonumber \\
\end{eqnarray}
with $|\cos{\theta_1}|=|\cos{\theta_{2}}|=|\cos{\theta_{3}}|=\cos{\theta}$. \\
The above expression can now be used to estimate the contribution of background anisotropy to the parameter $f_{NL}$. First, let us consider the ``flattened triangle'' limit, where $k_2\approx k_3 \approx k_1/2 \sim k$.\\
Naively,
\begin{equation}
\Delta f_{NL} \sim \frac{\Delta A({\bf k},{\bf k},{\bf k})}{P({\bf k})^2 }\approx \frac{\dot{\phi}^2}{\dot{\rho}^2}\left(\frac{\dot{\rho}}{k}\right)^{\frac{3}{2}} (k\eta_0)\left(-\frac{1}{3}+\cos^2{\theta}\right)\sin{\frac{2}{\varepsilon}}
\end{equation}
where $\left\langle \zeta_c({\bf k}_1,\eta)\zeta_c({\bf k}_2,\eta)\zeta_c({\bf k}_3,\eta)\right\rangle= \delta^{(3)}({\bf k}_1+{\bf k}_2+{\bf k}_3)A({\bf k}_1,{\bf k}_2,{\bf k}_3,\eta)$ and $\left\langle \zeta_{\bf k}(t)\zeta_{\bf k'}(t)\right\rangle= (2\pi)^3\delta^3({\bf k}+{\bf k'})P({\bf k})$.
However, as pointed out in \cite{Holman:2007na},  a factor of $|k\eta_0|$ is lost when one computes the $l$-space bi-spectrum. Therefore, the final contribution to $f_{NL}$ becomes,
 
\begin{equation}
\Delta f_{NL}=\epsilon \, \left(\frac{\dot{\rho}}{k}\right)^{\frac{3}{2}}\left(-\frac{1}{3}+\cos^2{\theta}\right)
\end{equation}
which, in addition to the standard slow-roll factor, is suppressed by powers of $ \frac{\dot{\rho}}{k}$, leading to an extremely small change in the value for $f_{NL}$.\\
Now, consider a ``squeezed''limit of this configuration: $k_3 \ll k_2 \approx k_1 \sim k$. In this case,
\begin{equation}
\Delta f_{NL} \sim \frac{\Delta A({\bf k},{\bf k},{\bf k_3})}{P({\bf k})P({\bf k_3}) }\approx \frac{\dot{\phi}^2}{\dot{\rho}^2}\left(\frac{\dot{\rho}}{k}\right)^{\frac{3}{2}} (k\eta_0)\left(-\frac{1}{3}+\cos^2{\theta}\right)
\end{equation} 
leading to the same final contribution to $f_{NL}$ as derived in the ``flattened triangle'' limit.\\
Equation (\ref{aRf}) can be studied in another interesting limit ,
namely $k_3\ll k_1 \approx k_2 \sim k$ but $|k_1-k_2|\neq k_3$ - usually known as the ``squeezed triangle'' limit. 
Note that the denominator $\sum_i\xi_ik_i=-k_3$, where
$|k_3\eta_0| \gg 1$ ensuring the mode is sub-horizon around the time when the
universe isotropizes.\\
The resultant contribution to the bi-spectrum is given as,
\begin{eqnarray}
\Delta A({\bf k_1},{\bf k_2},{\bf k_3})&\approx& \frac{\sum_{i<j}k^2_ik^2_j}{\dot{\phi}^2}\frac{1}{16\dot{\rho}^3}\prod^3_{i=1}\varepsilon_i^6 \left[ \left(-\frac{1}{3}+\cos^2{\theta_1}\right)\varepsilon_1^3 e^{\frac{-2i}{\varepsilon_1}}+  \left(-\frac{1}{3}+\cos^2{\theta_2}\right)\varepsilon_2^3 e^{\frac{-2i}{\varepsilon_2}}\right] \nonumber \\
&&\left(\frac{1-e^{-ik_3\eta_0}}{k_3}\right) + c.c.
\end{eqnarray} 
Setting $ k_1 \approx k_2 \sim k$ in the above equation, we have,
\begin{equation}
\Delta A({\bf k_1},{\bf k_2},{\bf k_3})\approx \frac{k^4}{\dot{\phi}^2\dot{\rho}^3}\left(\frac{\dot{\rho}}{k}\right)^6 \left(\frac{\dot{\rho}}{k_3}\right)^{9/2}(\cos^2{\theta_1}+\cos^2{\theta_2}-2/3)\frac{[\cos{\frac{2}{\varepsilon}}+\cos{(\frac{2}{\varepsilon}+k_3\eta_0)}]}{k_3}
\end{equation}
Therefore,
\begin{equation}
|\Delta f_{NL}| \sim \frac{\Delta A({\bf k},{\bf k},{\bf k_3})}{P({\bf k})P({\bf k_3}) }\approx \epsilon\left(\frac{\dot{\rho}}{k}\right)^{\frac{3}{2}}\frac{k}{k_3}
\end{equation}
which shows that the enhancement for the "squeezed triangle" limit is greater compared to the "flattened triangle" by a factor of
$k/k_3$. The result is in agreement with  the findings of \cite{Agullo:2010ws}, where the bi-spectrum in the "squeezed" limit was shown to be greater than that obtained in the "flattened" limit by this precise factor of $k/k_3$, for a generic (i.e. non-Bunch Davies) choice of the vacuum state. It was also argued in \cite{Agullo:2010ws} that $(k/k_3)_{max} \sim 200$, given the range
of scales measurable in the CMB. For $\epsilon \sim 10^{-2} $, $\dot{\rho}\sim H_{I} \approx 10^{-6} M_p$ and comoving
wave-number $k \approx 10^{-4}M_p$, we have $|\Delta f_{NL}| \approx  10^{-3}$, which shows that the "squeezed triangle" enhancement is too small to produce any observable non-gaussianity.\\

As explained in \cite{Holman:2007na}, in order to have a larger $f_{NL}$, one needs to include terms involving higher derivatives in the scalar field, which are compatible with the slow-roll conditions. As explained in \cite{Creminelli:2003iq}, the lowest dimensional operator that can be included is a dimension 8 operator $\frac{(\nabla\phi)^4}{M^4}$, where $M$ is the cut-off scale for the effective field theory for
inflation. As a next step, we compute the contribution of the background anisotropy once the higher derivative interaction terms are also included. The computation roughly follows the treatment in  \cite{Holman:2007na}.\\
The single-field theory,analyzed above, should now be modified by adding a term $\mathcal{L}_I=\sqrt{-g}\frac{\lambda}{8M^4}(\nabla\phi)^4$. On expanding the lagrangian around the classical solution to third order in perturbation, we have the following interaction Hamiltonian for the curvature perturbation:
\begin{equation}
{\cal H}_I=-\int d^3x a(\eta)\frac{\lambda \dot{\phi}^4}{2H^3M^4}\zeta'(\zeta'^2-(\partial_i\zeta)^2)  \label{asf}
\end{equation}
  
As explained in  \cite{Holman:2007na},  the three-point correlation function at the tree level (from the vertices above), computed for an excited state, is enhanced for a flattened triangle configuration, as we had observed in the case of a single-field theory without higher derivative interaction. In addition, we shall have an enhancement for a  "squeezed triangle" configuration as well.

From ($\ref{asf}$), one can directly compute the 3-point correlation function,
\begin{eqnarray}
A_R({\bf k_1},{\bf k_2},{\bf k_3})&=&i\int^{0}_{\eta_0}d\eta e^{\rho(\eta)} \frac{\lambda \dot{\phi}^4}{\dot{\rho}^3 M^4} [\prod^{3}_{i=1}\partial_{\eta}G_{{\bf k_i}}(0,\eta)\times (3!) \nonumber \\ &+& ((\vec{k_1}.\vec{k_2}) G_{{\bf k_1}}(0,\eta)G_{{\bf k_2}}(0,\eta)\partial_{\eta}G_{{\bf k_3}}(0,\eta) +perms)\times (2!) ]  + c.c. \label{aqf}
\end{eqnarray}
where the factors of $3!$ and $2!$ are the respective combinatorial factors for the two vertices.\\
As before, the leading terms due to anisotropy will appear in the integrand as coefficients of $e^{(\sum_i \xi_i k_i)\eta}$, with $\xi_1=1, \xi_2=-1,\xi_3=-1$ and its permutations. It is, therefore, sufficient to isolate the contribution proportional to $\exp{\sum_i \xi_i k_i}$ (with $\xi_i$s specified above) in the integrand for computing
enhancements in the bi-spectrum in the flattened as well as the squeezed limit.

From the definition of $\prod^{3}_{i=1}\partial_{\eta}G_{{\bf k_i}}(0,\eta)$, we have,
\begin{equation}
\prod^{3}_{i=1}\partial_{\eta}G_{{\bf k_i}}(0,\eta)=(\frac{\dot{\rho}}{\dot{\phi}})^6 \eta^3 \frac{(\prod^3_{i=1}k^2_i \varepsilon^6_i)}{16 \dot{\rho}^3}[e^{i(k_1-k_2-k_3)\eta}\varepsilon^3_1(-1/3 + \cos^2{\theta_1})e^{-2i/\varepsilon_1} +  perms.] 
\end{equation}
\begin{eqnarray}
G_{{\bf k_1}}G_{{\bf k_2}}\partial_{\eta}G_{{\bf k_3}}=\left(\frac{\dot{\rho}}{\dot{\phi}}\right)^6 k^2_3\eta \frac{\prod^3_{i=1} \varepsilon^6_i}{16 \dot{\rho}^3} & &\left[ e^{i(k_1-k_2-k_3)\eta}(1-ik_1\eta +ik_2\eta +k_1 k_2 \eta^2)\varepsilon^3_1 \right. \nonumber \\ & & \left.  (-1/3 + \cos^2{\theta_1})e^{-2i/\varepsilon_1} + perms.\right]
\end{eqnarray}
Plugging these back into ($\ref{aqf}$), we obtain
\begin{eqnarray}
A_R=i\int^{0}_{\eta_0}d\eta \frac{\lambda \dot{\phi}^4}{\dot{\rho}^4 M^4} \frac{(\prod^3_{i=1} \varepsilon^6_i)}{16 \dot{\rho}^3} \left(\frac{\dot{\rho}}{\dot{\phi}}\right)^6 & &\left[  e^{i(k_1-k_2-k_3)\eta}\varepsilon^3_1(-1/3 + \cos^2{\theta_1})e^{-2i/\varepsilon_1}\mathcal{A}({\bf k_1},{\bf k_2},{\bf k_3},\eta) \right. \nonumber \\ & & \left. + perms.\right] + c.c.   \label{apr}
\end{eqnarray}   
where we have used $e^{\rho}\eta = 1/\dot{\rho}$. The explicit form for the function $\mathcal{A}({\bf k_1},{\bf k_2},{\bf k_3},\eta)$ is given as,
\begin{eqnarray}
\mathcal{A}({\bf k_1},{\bf k_2},{\bf k_3},\eta)&=&6\eta^2(\prod_i k^2_i) + (k^2_3-k^2_1-k^2_1)(1-ik_1\eta + ik_2\eta +k_1k_2 \eta^2)k^2_3 \nonumber \\
&+& (k^2_1-k^2_2-k^2_3)(1+ik_2\eta + ik_3\eta -k_2k_3 \eta^2)k^2_1  \nonumber \\
&+& (k^2_2-k^2_1-k^2_3)(1-ik_1\eta + ik_3\eta +k_1k_3 \eta^2)k^2_2 
\end{eqnarray}

From the above general expression, we can now proceed to compute $f_{NL}$ in the limits of interest.\\
Consider first  the flattened triangle limit: $k_2=k_3=k=k_1/2$. The function $\mathcal{A}({\bf k_1},{\bf k_2},{\bf k_3},\eta)$ reduces to,
\begin{eqnarray}
\mathcal{A}({\bf k_1},{\bf k_2},{\bf k_3},\eta)&=&24k^6\eta^2 + ik\eta(4k^4+16k^4+4k^4)+(-4k^4+8k^4-4k^4) + k^2\eta^2(-8k^2-8k^2 -8k^2) \nonumber \\
&=& 24ik^5\eta
\end{eqnarray}
Integrating out $\eta$, we have,
\begin{equation}
A_R \approx \frac{\lambda \dot{\rho}^8}{\dot{\phi}^2 M^4} \frac{1}{k^4} \varepsilon^{3} \eta^2_0 \cos{2/\varepsilon} (-1/3 + \cos^2{\theta})
\end{equation}

Therefore, the ratio of the enhancement of the bi-spectrum in the flattened triangle configuration to the bi-spectrum computed in the standard Bunch-Davis vacuum is given as,
\begin{equation}
\frac{\Delta\left\langle\zeta(\vec{k_1},\eta_0)\zeta(\vec{k_2},\eta_0)\zeta(\vec{k_3},\eta_0)\right\rangle}{\left\langle\zeta(\vec{k_1},\eta_0)\zeta(\vec{k_2},\eta_0)\zeta(\vec{k_3},\eta_0)\right\rangle_{BD}}\approx \varepsilon^3\left(-\frac{1}{3}+\cos^2{\theta}\right)|k\eta_0|^2 
\end{equation}
where $\Delta\left\langle\zeta({\bf k}_1,\eta_0)\zeta({\bf k} _2,\eta_0)\zeta_c({\bf k}_3,\eta_0)\right\rangle$ denotes the leading contribution of background anisotropy to the bi-spectrum as before.\\
However, as in the previous case, one factor of $|k\eta_0|$ drops out from the $l$-space bi-spectrum and as a result, the $f_{NL}$ parameter for these configurations changes by,
\begin{equation}
|\Delta f_{NL} |\approx \frac{\dot{\phi_e}^2}{M^4}\varepsilon^3 |k\eta_0| \label{nge}
\end{equation}
where $\varepsilon=\sqrt{\frac{\dot{\rho}_e}{k}}$, $|k \eta_0|=\frac{k}{a(\eta_0)H(\eta_0)}\le \frac{M}{H_I}$ and $\dot{\phi_e}^2$ is related to the slow-roll parameter 
$\epsilon$ as $\dot{\phi_e}^2 \approx \dot{\rho_e}^2 M_p^2 \epsilon $ (here we have replaced the parameters $\dot{\phi}$ and $\dot{\rho}$ by their values at horizon exit, 
assuming that these remain essentially constant in the de Sitter phase).\\
This implies that,
\begin{equation}
|\Delta f_{NL}| \approx  \frac{H_I \, M_P^2 \, \epsilon \, \varepsilon^3}{ M^3} \label{ngef}
\end{equation}

To estimate the range of values for $f_{NL}$, we take $H_I \approx H_e \approx \frac{\sqrt{V}}{M_p} \approx H(\eta_0)\approx 10^{-6}M_p$, where $\eta_0\geq \eta_{iso}$. We define a number $n$ such that $n=\frac{t_0}{t_{iso}}$.\\
 Now, for the range of momenta we are interested in, we require $k e^{-\alpha(\tau_{*})}\leq M$, where $\tau_*$ corresponds to the time at which the early-time WKB solution is matched with the late-time de Sitter solution (see appendix) and is defined as $e^{\alpha(\tau_{*})}=\sqrt{\frac{k}{H_{*}}}$. Combining the two relations, we clearly have $k^{1/2}\leq \frac{M}{H^{1/2}_{*}}$. Taking $H_{*} \approx H_{I}$,
we find $\varepsilon =\sqrt{\frac{H_{I}}{k}} \geq \frac{H_{I}}{M}$.\\ 

Therefore, plugging in the above in equation (\ref{ngef}), we have,
\begin{eqnarray}
|\Delta f_{NL}|  &\approx& \frac{H_I}{M_p}\left(\frac{M_p}{M}\right)^3\epsilon \left(\frac{H_I}{M}\right)^3 \nonumber \\
&\approx& \epsilon\left(\frac{H_I}{M_p}\right)^4\left(\frac{M_p}{M}\right)^6
\end{eqnarray}
Now, substituting the values for the slow-roll parameter  $\epsilon \approx 10^{-2}$ and the scale of inflation $H_{I} \approx  10^{-6}M_p$, the above equation implies that a large $f_{NL}$ will require $M \sim 10^{-4}-10^{-5}M_p$. For example, $f_{NL}\sim 100$ if we choose $M =2.5 \times 10^{-5}M_p$.\\

Next, we consider a "squeezed triangle" limit which was not addressed in \cite{Holman:2007na}:  $k_3 \ll k_1\approx k_2 \approx k$, with $k_1 \neq k_2 + k_3$. In this limit, the function $\mathcal{A}({\bf k_1},{\bf k_2},{\bf k_3},\eta)$ reduces to,
\begin{eqnarray}
\mathcal{A}({\bf k_1},{\bf k_2},{\bf k_3},\eta)&=& 6k^4 k^2_3\eta^2 + (-2k^2)(1+k^2\eta^2)k^2_3  + (-k^2_3)(1+ik\eta -kk_3\eta^2)k^2  \nonumber \\ &+ &(-k^2_3)(1-ik\eta +kk_3\eta^2)k^2 \nonumber \\
&=& -4k^2k^2_3 + 4k^4\eta^2 k^2_3\\
\end{eqnarray}
Now, 

\begin{eqnarray}
\int^0_{\eta_0}d\eta e^{i(k_1-k_2-k_3)\eta}(-4k^2k^2_3+4k^4k^2_3\eta^2) & \approx&  \int^0_{\eta_0}d\eta e^{-ik_3\eta}(-4k^2k^2_3+4k^4k^2_3\eta^2) \nonumber \\& &\sim   -i4k^3\left(\frac{k_3}{k}\right) -i4k^3 \left(\frac{k}{k_3}\right) \sim -i4k^3 \left(\frac{k}{k_3}\right) \nonumber \\
\end{eqnarray} since $k \gg k_3$. In evaluating the integral we have assumed that $|k_3\eta| \gg 1$ for $\eta \in [\eta_0,0]$.

From (\ref{apr}), one obtains the following bispectrum,
\begin{equation}
A_R \approx \frac{\lambda \dot{\rho}^8}{\dot{\phi}^2 M^4} \frac{1}{k^2 k^4_3} \varepsilon^{3}  \cos{2/\varepsilon} (-2/3 + \cos^2{\theta}_1 +\cos^2{\theta}_2)
\end{equation}
where $\varepsilon=\sqrt{\frac{\dot{\rho}}{k}}$. \\
The corresponding $f_{NL}$ can be read off as follows:
\begin{equation}
|\Delta f_{NL}|= \frac{\dot{\phi_e}^2}{M^4}\varepsilon^3 \left(\frac{k}{k_3}\right)=\epsilon\left(\frac{H_I}{M_p}\right)^5\left(\frac{M_p}{M}\right)^7 \frac{k}{k_3} \label{pdr}
\end{equation}
Evidently, 
\begin{equation}
\frac{|\Delta f^{flat}_{NL}|}{|\Delta f^{squeezed}_{NL}|}= |k_3\eta_0| \gg 1
\end{equation}
which shows that enhancement in the bi-spectrum for flattened triangle limit can be more important than squeezed triangle limit, once we include higher-derivative terms
in the interaction Hamiltonian.\\
To estimate the range of $f^{squeezed}_{NL}$, we need to ascertain the maximum value of $\frac{k}{k_3}$. Recall that $|k\eta_0| \leq \frac{M}{H_I}$. Therefore, taking
$M \sim 10^{-4} M_p$, we may set $|k\eta_0|_{max} \sim 100$. Now, since $|k_3\eta_0| \gg 1 $, one can also set $|k_3\eta_0|_{min} \sim 10 $. This naturally fixes  
the ratio $(\frac{k}{k_3})_{max} \sim 10$ - one order lower than the estimated value in \cite{Agullo:2010ws}. In the squeezed limit, therefore, the scale $M$ cannot be lowered
much below $10^{-4} M_p$, as it pushes the ratio $(\frac{k}{k_3})_{max}$ closer to $O(1)$. Plugging these values in (\ref{pdr}) we have, 
$|\Delta f^{squeezed}_{NL}|\approx 10^{-3}$, which is too small compared to the observable limits of non-gaussianity.\\
This proves that the flattened triangle is the dominant source of non-Gaussianity for the curvature perturbation spectrum given an interaction Hamiltonian of the form (\ref{asf}).\\

Finally, the particular value of the cut-off scale $M$ for the effective field theory merits some explanation. In \cite{Weinberg:2008hq}, it was argued that the scale $M$ cannot be too small compared to $\sqrt{2\epsilon}M_p$ if one wishes to impose a limit on the size of higher derivative terms in the effective action. For $M =5 \times 10^{-4}M_p$, however, one cannot rule out the presence of higher dimensional operators in the effective action and their contributions to $\Delta f_{NL}$. These contributions will be suppressed by factors of $\frac{H_I}{M}$ \cite{Creminelli:2003iq} where  $\frac{H_I}{M} \sim 10^{-2}-10^{-1}$ in our case (as opposed to a suppression factor of $10^{-5}$ in  \cite{Weinberg:2008hq})\\

\section{Conclusion}
In this note, we have presented a way to probe the signature of pre-inflationary background anisotropy in the spectrum of cosmological perturbations at late times for a range of high-momentum non-planar modes. These modes, which generically exit the horizon after the universe has isotropized, have a nice  WKB description at early times. The late time dynamics of these modes is characterized by an excited state built on the standard Bunch-Davies vacuum and this state carries the signature of the pre-inflationary anisotropy. We have computed correlation functions of the curvature perturbation (two-point and three-point functions) in  this new ground state and investigated the issue of having possible  observable signatures of anisotropy.\\

Our computation suggests that contribution of anisotropy to the spectral index could be appreciable even for the range of large momenta modes (for which the WKB approximation is valid) which exit the horizon after isotropization. However, if the number $n=\frac{t_e}{t_{iso}}$ is of the order of 10, any correction to the two-point function is severely suppressed in a single-field inflationary model with a canonical action. The effect of anisotropy will obviously be much larger for modes with small wavenumbers which exit the horizon before isotropization. The two-point function in this regime was numerically analyzed in  \cite{Gumrukcuoglu:2007bx}.\\

The three-point function for the curvature perturbation, in this case, is enhanced in a flattened triangle as well as a squeezed triangle limit, although the latter leads to a larger $f_{NL}$ .  However, the bispectrum in either case is suppressed by factors of $\frac{H_e}{k}$ in addition to the usual suppression by the slow-roll parameter, making the contribution negligible compared to observable limits. \\

The problem can be circumvented, to a certain extent, by including higher derivative operators in the action and in this work, 
we have studied the effect of having a dimension 8 operator $\frac{\lambda}{8M^4}(\nabla\phi)^4$ in the action. Computation of the resultant three-point correlation function shows that the bi-spectrum for the curvature perturbation is again enhanced for a flattened triangle configuration and a squeezed triangle configuration, with the former
being the dominant source of non-Gaussianity  in this case - $\frac{|\Delta f^{flat}_{NL}|}{|\Delta f^{squeezed}_{NL}|} \sim 10$ for the cut-off scale $M \sim 10^{-4} M_p$.\\

 We have shown that the $f_{NL}$ for the squeezed triangle case will always be extremely small, since, for reasons explained at the end of the previous section, one cannot
 lower the cut-off scale much beyond $10^{-4} M_p$. For a large $f_{NL} \approx 100$, in the flattened triangle case, one needs to have the cut-off scale for the effective field theory to be set around $M \approx 10^{-5}M_p$, which is low enough for higher-dimensional operators (suppressed by factors of $\frac{H_I}{M}$) to appear in the effective action. We observe that this result is very similar to that obtained in \cite{Holman:2007na} where the authors studied the enhancement of the bi-spectrum for an excited state in a single-field theory with a dimension 8 operator. \\
In our case, however, the effective ground state is characterized by an angular dependence and a particular momentum dependence ($\sim (\sqrt{\frac{\dot{\rho}}{k}})^3$). These are the distinctive signatures of background anisotropy  in the enhanced non-Gaussianity  for  the flattened triangle and the squeezed triangle configurations, distinguishing it from a generic case of enhanced bi-spectrum for a non-BD vacuum state.\\
Finally, we want to remind the reader that this analysis holds only for the non-planar modes, i.e.  modes with $k_1 \approx k_2, k_3$ for which the WKB condition always holds at early times. The planar case and non-Gaussianities from the interplay of these two regimes are the subjects of a work in progress.\\

\section{Acknowledgements}

We would like to thank M. Peloso and E. Komatsu for valuable discussions and Iv\'{a}n Agull\'{o} for useful comments and suggestions. 

This research was supported in part by the National Science Foundation under Grant Numbers PHY-0969020 and PHY-0455649

\section{Appendix: Scalar Field in Anisotropic Background}
In this section, we study the evolution of a single, massless scalar field minimally coupled to gravity in an axially symmetric anisotropic space-time, with a positive cosmological constant. In particular, we construct a WKB solution for the scalar field at early times and show how one can match it with the general solution at late times to obtain equations (\ref{apam}). The treatment essentially follows \cite{Kim:2010wra}\\
We consider the following action for the scalar field,
\begin{equation}
S= - \int d^4x \sqrt{-g}\left(\frac{1}{2}g^{\mu\nu}\partial_{\mu}\phi\partial_{\nu}\phi+ V\right), \hspace{4ex} (M_p^2\equiv 1)
\end{equation}
where the background metric is chosen to be an axially symmetric version of the
Bianchi I metric:
\begin{equation}
ds^2=-dt^2 + \exp{(2\rho)}(dx^1)^2 + \exp{(2\beta)}(dx^{\alpha})^2, \hspace{4ex}(\alpha=2,3)
\end{equation}
where  $\rho,\beta$ are known functions of time:
\begin{eqnarray}
%\rho& =& \ln{\frac{\sinh{\frac{3H_I t}{2}}}{\cosh^{\frac{1}{3}}{3H_I t}}}\\ \nonumber \\
%\beta& =&\frac{2}{3}  \ln{\cosh{\frac{3H_I t}{2}}}
\rho& =& \frac{1}{3}\ln{\tanh^{2}{\left(\frac{3H_I t}{2}\right)}\sinh{(3H_I t)}} \nonumber \\
\beta& =& \frac{1}{3}\ln{\left[\frac{\sinh{(3H_I t)}}{\tanh{\left(\frac{3H_I t}{2}\right)}}\right]}
\end{eqnarray}
with $H_I=\sqrt{\frac{V}{3}}$.\\
Define $\rho=\alpha - 2\beta_+, \beta=\alpha+\beta_+$ and a new ``time'' coordinate $\tau$, analogous to the conformal time in the isotropic limit,  as,
\begin{equation}
d\tau =\frac{dt}{e^{3 \alpha}}
\end{equation}

From equation (7.3), one can derive, $e^{3\alpha}= e^{\rho+2\beta}=\sinh{(3H_I t)}= \frac{1}{\sinh{(-3H_I \tau)}}$.\\
It can be easily seen that as $t$ varies from $0^+$ to $\infty$, $\tau$ varies from $-\infty$ to $0^-$. In this time coordinate, the equation of motion for a mode $\phi_k$ is given by,
\begin{equation}
\left(\frac{d^2}{d\tau^2}+\omega(\tau)^2\right)\phi_k=0
\end{equation}
The frequency squared is given as,
\begin{equation}
\omega(\tau)^2=\frac{2^{\frac{4}{3}}k^2}{x^{\frac{4}{3}}}(1-r^2 x)
\end{equation}
where $r^2=\frac{k_2^2+k_3^2}{k^2}$ and $x(\tau)=1-e^{6H_{I}\tau}=\exp{(-6\alpha)}(\sqrt{\exp{(6\alpha)}+1}-1)$. 
Evidently, $x(\tau)$ varies from 1 to 0 as $\tau$ changes from $-\infty$ to $0$.\\
Equation (7.6) has a WKB solution:
\begin{equation}
\phi_{WKB}=\frac{1}{\sqrt{2\tilde{\omega}}}\exp{[-i\int^{\tau}_{\tau_0}d\tau^{'}
\tilde{\omega}]}
\end{equation}
where $\tilde{\omega}$ has to be determined from the equation,
\begin{equation}
\tilde{\omega}^2=\omega^2 -\frac{1}{2}\left(\frac{\tilde{\omega}_{,\tau\tau}}{\tilde{\omega}}-\frac{3\tilde{\omega}^2_{,\tau}}{2\tilde{\omega}^2}\right)
\end{equation}
The WKB approximation holds as long as the WKB parameter
\begin{eqnarray}
\varepsilon &= & \left|\frac{\frac{d\omega^2}{d\tau}}{\omega^3}\right| \nonumber \\ \nonumber \\
& = & \frac{H_{I}}{k}\frac{1-x(\tau)}{(x/2)^{1/3}(1-r^2x(\tau))^{1/2}}\left(\frac{3}{1-r^2x(\tau)}+1\right)\ll 1
\end{eqnarray} 
The choice of the WKB solution above is obviously equivalent to imposing a particular initial condition on the 
modes of the scalar field at early times. This can be seen directly by analyzing the early time behavior of the classical solution for the scalar field. Firstly note that $\tilde{\omega} \approx \omega \to k_1$ in the limit $\tau\to -\infty$ (or $t \to 0+$) and in this limit $\tau$ and $t$ are related as $\tau=\frac{1}{3H_I} \ln{\frac{3H_I t}{2}}$. Therefore, in the early time limit, the time-dependence of the WKB solution is given as follows,
\begin{equation}
\phi_{WKB} \approx \frac{1}{\sqrt{2 k_1}} \exp[-i\frac{k_1}{3H_I} \log{\frac{3H_I t}{2}}] \sim t^{-i\frac{k_1}{3H_I} }
\end{equation}
Now consider the equation of motion of the scalar field at early times,
\begin{equation}
\ddot{\phi_k} +\frac{1}{t}\dot{\phi_k} + (k_2^2 +k_3^2 + \frac{k_1^2}{3Vt^2})\phi_k=0 \label{earlytime}
\end{equation}
Define $z= \ln{\sqrt{k_2^2 +k_3^2 } t}$, so that the equation reduces to 
\begin{equation}
\phi^{''}_k(z) + (e^{2z} +\frac{k_1^2}{3V} )\phi_k(z)=0
\end{equation}
In the limit $t \to 0+$, $z \to -\infty$, so that the exponential term drops out of the above equation and we obtain a solution of the form, 
\begin{equation}
\phi_k(z)= A(k) e^{-i\frac{k_1}{\sqrt{3V}}z } + B(k) e^{i\frac{k_1}{\sqrt{3V}}z }
\end{equation}
Choosing $B(k)=0$, we find that $\phi_k(z) \sim t^{-i\frac{k_1}{3H_I} }$, confirming that the WKB solution has the same time-dependence at early times as expected from the classical solution subject to a certain initial condition.\\
In fact, equation (\ref{earlytime}) has the general solution, 
\begin{equation}
\phi_k(t)= C_1(k) H^{(1)}_{ik_1/\sqrt{3V}} (\sqrt{k_2^2 +k_3^2 } t) + C_2(k) H^{(2)}_{ik_1/\sqrt{3V}} (\sqrt{k_2^2 +k_3^2 } t)  
\end{equation}
Using $H^{(1)}_{i\nu} (z)=\frac{1}{\sinh{\pi \nu}} (J_{i\nu}(z) e^{\pi \nu} - J_{-i\nu}(z))$ and $H^{(2)}_{i\nu} (z)=\frac{1}{\sinh{\pi \nu}} (-J_{i\nu}(z) e^{-\pi \nu}+ J_{-i\nu}(z))$, we can rewrite the general solution as,
\begin{equation}
\phi_k(t)=A(k) J_{ik_1/\sqrt{3V}}(\sqrt{k_2^2 +k_3^2 } t) + B(k) J_{-ik_1/\sqrt{3V}}(\sqrt{k_2^2 +k_3^2 } t)
\end{equation}
Therefore, the WKB solution corresponds to imposing the initial condition $A(k)=0$ and choosing $B(k)$ appropriately, as $J_{-ik_1/\sqrt{3V}}(\sqrt{k_2^2 +k_3^2 } t)$ has the same time-dependence at early times as our WKB solution.\\

We will be interested in the large momentum regime of non-planar wavenumbers, i.e. $k_i\gg H$, implying that the factor $(1-r^2x(\tau))$ in the denominator doesn't vanish anywhere (since both $r$ and $x$ are fractions). In the regime $x\approx 1$ (i.e. early times), the WKB condition obviously holds for any momentum. In \cite{Kim:2010wra}, it was shown that the condition holds  for high-momentum modes as long as  $\frac{k}{H} \gg \exp{\alpha(t)}$. Therefore, the time at which the WKB solution should be matched with the late-time de Sitter solution has a natural choice, $\tau_{*}$, such that,
\begin{equation}
e^{\alpha(\tau_{*})}=\sqrt{\frac{k}{H_I}}
\end{equation}
The above equation implicitly states that $\tau_{*}$ corresponds to late times when $e^{\alpha(\tau_*)} \gg 1$. In terms of real time, this condition implies that $e^{\alpha(t_*)} \approx e^{H_I t_*} \gg 1$, or $t_* > t_{iso} \sim \frac{1}{H_I}$. Also, if $t_e$ denotes the time of horizon exit for a given mode of wavenumber $k$, we have $e^{H_I t_e} =\frac{k}{H}$. This suggests a simple relation between $t_*$ and $t_e$, viz.
\begin{equation}
t_e \approx 2 t_*
\end{equation}

Returning to the problem of matching the modes - since $k\gg H_I$, $x(\tau_*) \approx \exp{(-6\alpha(\tau_*))} \approx 0$. Therefore the WKB solution around $\tau=\tau_*$ is given by expanding equation (7.7) around $x=0$ and then plugging in the values of $x$ and the frequencies at $t=t_*$.
The solution can be expanded in powers of $\varepsilon=\sqrt{\frac{H_I}{k}}$ and one needs to retain terms to the order at which the direction dependence first appears. It turns out that it is sufficient to retain terms up to the order $\varepsilon^3$ and to this order $x(\tau_*)$ and the frequencies are given as,
\begin{equation}
x(t_*)=2(H_I/k)^{3/2}(1-(H_I/k)^{3/2})
\end{equation}
\begin{equation}
\omega_*=\frac{k^2}{H_I} \, \left[1+(\frac{2}{3}-r^2)(\frac{H_I}{k})^{3/2}\right]
\end{equation}
\begin{equation}
\tilde{\omega}_*^2=\omega_*^2 \left(1-2\frac{H_I}{k}\right)
\end{equation}

Now, in the de Sitter regime the solution to equation (7.5) is given as,
\begin{equation}
\phi_k=A_+\phi_+(\tau) + A_-\phi_-(\tau)
\end{equation}
where the modes $\phi_{\pm}$ are given as,
\begin{equation}
\phi_{\pm}(\tau)=\left(1 \mp \frac{ik}{H_I}(-3H_I\tau)^{1/3}\right)\exp{\left(\pm \frac{ik}{H_I}(-3H_I\tau)^{1/3}\right)}
\end{equation}
Matching the de Sitter solution with the WKB solution at $\tau=\tau_*$, we obtain the following equations for $A_+$ and $A_{-}$,
\begin{eqnarray}
A_+\phi'_+(\tau_*) + A_-\phi'_-(\tau_*) & = & -i\left[\left(1-\frac{H_I}{k}\right)-i\sqrt{\frac{H_I}{k}}\left(1-\frac{H_I}{2k}\right)\right]\sqrt{\frac{\omega(t_*)}{2}}\\
A_+\phi_+(\tau_*) + A_-\phi_-(\tau_*)& = & \left(1+\frac{H_I}{2k}\right)\sqrt{\frac{1}{2\omega_*}}
\end{eqnarray}
where we have absorbed an overall phase in the definition of $A_+$
and $A_-$. Solving for $A_{\pm}$ from the above equations, we have,
\begin{eqnarray}
A_{+} & = & \frac{i\varepsilon^3}{2\sqrt{2H_I}}\left[\left(2-\varepsilon^2 \right)+2i\varepsilon\left(\frac{\varepsilon^2}{2}-1\right) +O(\varepsilon^4)\right]\exp{\left(\frac{-i}{\varepsilon}\right)}\\
A_{-}& = & \frac{i\varepsilon^3}{2\sqrt{2H_I}}\left[\left(\frac{2}{3}-r^2 \right)\varepsilon^3 +O(\varepsilon^4) \right]\exp{\left(\frac{i}{\varepsilon}\right)}
\end{eqnarray}
One can easily verify that these coefficients obey the normalization condition,
\begin{equation}
|A_+|^2-|A_-|^2=\frac{H_I^2}{2k^3}
\end{equation}


\newpage

\begin{thebibliography}{19} 

%\cite{Komatsu:2009kd}
\bibitem{Komatsu:2009kd}
  E.~Komatsu, N.~Afshordi, N.~Bartolo, D.~Baumann, J.~R.~Bond, E.~I.~Buchbinder, C.~T.~Byrnes, X.~Chen {\it et al.},
 ``non-Gaussianity as a Probe of the Physics of the Primordial Universe and the Astrophysics of the Low Redshift Universe,''
   [arXiv:0902.4759 [astro-ph.CO]].
   
   %\cite{Maldacena:2002vr}
\bibitem{Maldacena:2002vr}
  J.~M.~Maldacena,
  ``Non-Gaussian features of primordial fluctuations in single field inflationary models,''
  JHEP {\bf 0305}, 013 (2003)
  [arXiv:astro-ph/0210603].
  %%CITATION = JHEPA,0305,013;%%
   
   %\cite{Creminelli:2004yq}
\bibitem{Creminelli:2004yq}
  P.~Creminelli, M.~Zaldarriaga,
  %``Single field consistency relation for the 3-point function,''
  JCAP {\bf 0410}, 006 (2004).
  [astro-ph/0407059].
  
  %\cite{Bartolo:2004if}
\bibitem{Bartolo:2004if}
  N.~Bartolo, E.~Komatsu, S.~Matarrese, A.~Riotto,
 ``non-Gaussianity from inflation: Theory and observations,''
  Phys.\ Rept.\  {\bf 402}, 103-266 (2004).
  [astro-ph/0406398];
  X.~Chen,
``Primordial Non-Gaussianities from Inflation Models,''
  Adv.\ Astron.\  {\bf 2010}, 638979 (2010).
  [arXiv:1002.1416 [astro-ph.CO]];D.~Wands,
  %``Local non-Gaussianity from inflation,''
  Class.\ Quant.\ Grav.\  {\bf 27}, 124002 (2010).
  [arXiv:1004.0818 [astro-ph.CO]].
  
%\cite{Holman:2007na}
\bibitem{Holman:2007na}
  R.~Holman and A.~J.~Tolley,
``Enhanced non-Gaussianity from Excited Initial States,''
  JCAP {\bf 0805}, 001 (2008)
  [arXiv:0710.1302 [hep-th]].
  %%CITATION = JCAPA,0805,001;%%

%\cite{Gumrukcuoglu:2007bx}
\bibitem{Gumrukcuoglu:2007bx}
  A.~E.~Gumrukcuoglu, C.~R.~Contaldi and M.~Peloso,
  ``Inflationary perturbations in anisotropic backgrounds and their imprint on the CMB,''
  JCAP {\bf 0711}, 005 (2007)
  [arXiv:0707.4179 [astro-ph]].
  %%CITATION = JCAPA,0711,005;%%

%\cite{Creminelli:2003iq}
\bibitem{Creminelli:2003iq}
  P.~Creminelli,
  ``On non-Gaussianities in single-field inflation,''
  JCAP {\bf 0310}, 003 (2003)
  [arXiv:astro-ph/0306122].
  %%CITATION = JCAPA,0310,003;%%

%\cite{Gumrukcuoglu:2007bx}
\bibitem{Gumrukcuoglu:2007bx}
  A.~E.~Gumrukcuoglu, C.~R.~Contaldi, M.~Peloso,
  `Inflationary perturbations in anisotropic backgrounds and their imprint on the CMB,''
  JCAP {\bf 0711}, 005 (2007).
  [arXiv:0707.4179 [astro-ph]].; 
  
  %\cite{Pereira:2007yy}
\bibitem{Pereira:2007yy}
  T.~S.~Pereira, C.~Pitrou, J.~-P.~Uzan,
  `Theory of cosmological perturbations in an anisotropic universe,''
  JCAP {\bf 0709}, 006 (2007).
  [arXiv:0707.0736 [astro-ph]];  C.~Pitrou, T.~S.~Pereira and J.~P.~Uzan,
  ``Predictions from an anisotropic inflationary era,''
  JCAP {\bf 0804}, 004 (2008)
  [arXiv:0801.3596 [astro-ph]];
  

%\cite{Kim:2010wra}
\bibitem{Kim:2010wra}
  H.~C.~Kim and M.~Minamitsuji,
 ``Scalar field in the anisotropic universe,''
  Phys.\ Rev.\  D {\bf 81}, 083517 (2010)
  [Erratum-ibid.\  D {\bf 82}, 109904 (2010)]
  [arXiv:1002.1361 [gr-qc]].
  %%CITATION = PHRVA,D81,083517;%%

%\cite{Kim:2011pt}
\bibitem{Kim:2011pt}
  H.~C.~Kim and M.~Minamitsuji,
``An Analytic approach to perturbations from an initially anisotropic universe,''
  JCAP {\bf 1103}, 038 (2011)
  [arXiv:1101.0329 [gr-qc]].
  %%CITATION = JCAPA,1103,038;%%

%\cite{Gumrukcuoglu:2008gi}
\bibitem{Gumrukcuoglu:2008gi}
  A.~E.~Gumrukcuoglu, L.~Kofman and M.~Peloso,
``Gravity Waves Signatures from Anisotropic pre-Inflation,''
  Phys.\ Rev.\  D {\bf 78}, 103525 (2008)
  [arXiv:0807.1335 [astro-ph]].
  %%CITATION = PHRVA,D78,103525;%%


%\cite{Giblin:2010bd}
\bibitem{Giblin:2010bd}
  J.~T.~Giblin, Jr, L.~Hui, E.~A.~Lim, I-S.~Yang,
  %``How to Run Through Walls: Dynamics of Bubble and Soliton Collisions,''
  Phys.\ Rev.\  {\bf D82}, 045019 (2010).
  [arXiv:1005.3493 [hep-th]].

%\cite{Weinberg:2008hq}
\bibitem{Weinberg:2008hq}
  S.~Weinberg,
  %``Effective Field Theory for Inflation,''
  Phys.\ Rev.\  D {\bf 77}, 123541 (2008)
  [arXiv:0804.4291 [hep-th]].
  %%CITATION = PHRVA,D77,123541;%%

%\cite{Ganc:2011dy}
\bibitem{Ganc:2011dy}
  J.~Ganc,
  %``Calculating the local-type fNL for slow-roll inflation with a non-vacuum
  %initial state,''
  arXiv:1104.0244 [astro-ph.CO].
  %%CITATION = ARXIV:1104.0244;%%
  
  %\cite{Agullo:2010ws}
\bibitem{Agullo:2010ws}
  I.~Agullo and L.~Parker,
  %``Non-gaussianities and the Stimulated creation of quanta in the inflationary
  %universe,''
  Phys.\ Rev.\  D {\bf 83}, 063526 (2011)
  [arXiv:1010.5766 [astro-ph.CO]].
  %%CITATION = PHRVA,D83,063526;%%

%\cite{Yokoyama:2008xw}
\bibitem{Yokoyama:2008xw}
  S.~Yokoyama and J.~Soda,
  %``Primordial statistical anisotropy generated at the end of inflation,''
  JCAP {\bf 0808}, 005 (2008)
  [arXiv:0805.4265 [astro-ph]].
  %%CITATION = JCAPA,0808,005;%%
  
  %\cite{Qiu:2010dk}
\bibitem{Qiu:2010dk}
  T.~Qiu and K.~C.~Yang,
  %``Non-Gaussianities of Single Field Inflation with Non-minimal Coupling,''
  Phys.\ Rev.\  D {\bf 83}, 084022 (2011)
  [arXiv:1012.1697 [hep-th]].
  %%CITATION = PHRVA,D83,084022;%%
  
  %\cite{BlancoPillado:2010uw}
\bibitem{BlancoPillado:2010uw}
  J.~J.~Blanco-Pillado and M.~P.~Salem,
  %``Observable effects of anisotropic bubble nucleation,''
  JCAP {\bf 1007}, 007 (2010)
  [arXiv:1003.0663 [hep-th]].
  %%CITATION = JCAPA,1007,007;%%
  
  %\cite{Chialva:2011iz}
\bibitem{Chialva:2011iz}
  D.~Chialva,
  %``Enhanced CMBR non-Gaussianities from Lorentz violation,''
  arXiv:1106.0040 [hep-th].
  %%CITATION = ARXIV:1106.0040;%%
  
  %\cite{Ackerman:2007nb}
\bibitem{Ackerman:2007nb}
  L.~Ackerman, S.~M.~Carroll and M.~B.~Wise,
  %``Imprints of a Primordial Preferred Direction on the Microwave Background,''
  Phys.\ Rev.\  D {\bf 75}, 083502 (2007)
  [Erratum-ibid.\  D {\bf 80}, 069901 (2009)]
  [arXiv:astro-ph/0701357].
  %%CITATION = PHRVA,D75,083502;%%

\end{thebibliography}
\end{document}